\documentclass[aps,prd,amssymb,amsmath,amsfonts,superscriptaddress,nofootinbib,eqsecnum,reprint,showpacs,longbibliography]{revtex4-2}
\usepackage{graphicx}
\usepackage{dcolumn}
\usepackage{bm}

\usepackage{color}
\usepackage{xcolor}
\usepackage{float}
\usepackage{url}
\usepackage{subfigure}

\usepackage{url}
\usepackage{subfigure}
\usepackage{enumitem}
\usepackage{rotating}
\usepackage{multirow}
\usepackage{aas_macros}
\definecolor{linkcolor}{rgb}{0.0,0.3,0.5}
\usepackage[unicode,colorlinks=true,citecolor=linkcolor,linkcolor=linkcolor,urlcolor=linkcolor]{hyperref}

\usepackage{cleveref}

\graphicspath{{../figures/}}

\newcommand{\Ms}{M_{\odot}}

\newcommand{\Bplus}{Bauswein+\ }
\newcommand{\Kplus}{K\"{o}lsch+\ }
\newcommand{\PhenomDNRT}{\texttt{IMRPhenomD\_NRTidal}}
\newcommand{\PhenomPNRT}{\texttt{IMRPhenomPv2\_NRTidal}}
\newcommand{\TF}{\texttt{TaylorF2}}
\newcommand{\SEOBNRT}{\texttt{SEOBNRv4\_ROM\_NRTidal}}
\newcolumntype{F}[1]{%
    >{\raggedright\arraybackslash\hspace{0pt}}p{#1}}%
\newcolumntype{T}[1]{%
    >{\centering\arraybackslash\hspace{0pt}}p{#1}}%
 
\begin{document}
\preprint{APS/123-QED}
\title{Improving inference on neutron star properties using information from binary merger remnants}

\author{Tamanna Jain}%
 \email{tj317@cam.ac.uk}
\affiliation{%
Department of Applied Mathematics and Theoretical Physics, University of Cambridge, Wilberforce Road CB3 0WA Cambridge, United Kingdom.
}%

\author{Michalis Agathos}
\email{m.agathos@qmul.ac.uk}
\affiliation{
Queen Mary University of London, Mile End Road, London, E1
4NS, UK.\\
}
\affiliation{Department of Applied Mathematics and Theoretical Physics, University of Cambridge, Wilberforce Road CB3 0WA Cambridge, United Kingdom.
}


\date{\today}

\begin{abstract}
The gravitational-wave signal GW170817 is a result of a binary neutron star coalescence event.  The observations of electromagnetic counterparts suggest that the event didn't led to the prompt formation of a black-hole.  In this work, we first classify the GW170817 LIGO-Virgo data sample into prompt collapse to a black-hole using the $q$-dependent threshold mass fits and then remove these cases from the data sample. We find that the cases without a prompt black-hole formation do not support radii $ <$ 10~km unlike the LIGO-Virgo data sample. This is consistent with the maximum mass constraint, based on the binary pulsar J0348+0432, imposed LIGO-Virgo data sample. Additionally, we find that the cases without the prompt collapse to a black-hole improve the uncertainty range of neutron star radii from 3.3~km to 2.6~km for the data sample without the mass constraint and from 2.8~km to 2.5~km for the data sample with the mass constraint, implying improved constraints on the neutron star radii and hence the equation-of-state.
\end{abstract}

\maketitle


\section{\label{sec:intro}Introduction}
In 2015, the first gravitational-wave (GW) detected from the merger of a binary black-hole (BBH) \cite{LIGOScientific:2016aoc} by the LIGO-Virgo detectors \cite{LIGOScientific:2014pky,VIRGO:2014yos} opened a new era of GW astronomy.
Subsequently, LIGO-Virgo detected GW signals from coalescence of binary neutron-star (BNS) systems, namely GW170817 \cite{LIGOScientific:2017vwq,LIGOScientific:2018hze,LIGOScientific:2018mvr} and GW190425 \cite{LIGOScientific:2020aai}.

The merger of a BNS system can lead either to a prompt collapse to a BH or a formation of a NS remnant (non-prompt collapse) \cite{Shibata:2005xz,Shibata:2006nm,Baiotti:2008ra,Hotokezaka:2011dh,Bauswein:2013jpa,Bauswein:2017aur,Bernuzzi:2020tgt}, that can be either stable or unstable, depending on its mass and spin.
The detection of electromagnetic (EM) counterparts from space-based or ground-based observatories is crucial to identify the signal from the BNS merger as a non-prompt collapse signal. The EM radiation emitted can either be from the dynamical ejection of neutron-star matter, or from the formation of a disk surrounding the remnant.
The follow-up searches after the detection of the GW signal led to the observation of EM counterparts in the case of GW170817, while in the case of GW190425 no EM signal was detected.
The EM counterpart of GW170817, known as AT2017gfo, consisted of a short gamma ray burst (sGRB) \cite{LIGOScientific:2017zic,Troja:2017nqp,LIGOScientific:2017ync} followed up by strong emission from X-ray to radio  \cite{Tanaka:2017qxj,Tanvir:2017pws,Cowperthwaite:2017dyu,Nicholl:2017ahq,Chornock:2017sdf,Coulter:2017wya}, that was compatible with a kilonova transient event, and pointed to a non-prompt collapse merger.
There could be many reasons for the lack of detection of the EM counterparts for GW190425. Due to its poor sky-localization ($8284 \deg^2$ as its 90\% confidence region, compared to $28 \deg^2$ for GW170817), and its large distance ($159^{+69}_{71}\mathrm{Mpc}$ compared to $\sim40 \mathrm{Mpc}$ for GW170817), it was very unlikely that a low-latency survey would have picked up such a faint source while having to cover such a wide patch.
It is also possible that the EM emission was highly collimated and that our line of sight lied outside the source's emission cone.

Over the years, various numerical relativity (NR) studies investigated the remnants of BNS mergers, concluding that the coalescence of a BNS system is followed by the prompt collapse to a BH if the total mass of the binary is greater than some characteristic mass called the threshold mass.
Thus, accurate modelling of this threshold mass is important as it characterizes the outcome of the merger remnant for a BNS of a given total mass. The first use of the non-prompt-collapse hypothesis for GW170817 was made in Ref.~\cite{Bauswein:2017vtn} to place a bound on the NS radius, $R_{1.6} > 10 \, \mathrm{km}$, for a fiducial NS mass of $1.6 M_\odot$.

In recent works, Bauswein \textit{et al.}~\cite{Bauswein:2020xlt} and K\"olsch \textit{et al.}~\cite{Kolsch:2021lub} provide empirical formulae for the threshold mass, as a function of the equation of state (EOS) of NS matter and the mass ratio of the BNS, based on NR simulations.
In this work we make use of these fits and the presence of an EM counterpart for GW170817 event to improve the inference on the properties of NS matter. 
On applying this method to GW170817 we find that a significant part of the parameter space is not supported by the observation of the EM counterpart.

The paper is organised as follows.
In Sec.~\ref{NumSim}, we briefly discuss NR simulation techniques used in  Refs.~\cite{Bauswein:2020xlt,Kolsch:2021lub,Agathos:2019sah} and the range of EOSs and mass-ratios $q$ used in these studies.
Section \ref{sec:methods} first outlines the method based on which we classify the outcome of the BNS merger and then the Bayseian analysis used for GW170817.
Finally, we present our results for the GW170817 signal, along with the results from the injection studies of Ref.~\cite{Agathos:2019sah} in Sec.~\ref{sec:results} with concluding remarks in Sec.~\ref{sec:conclusions}.

\section{Overview of Numerical Simulations of Neutron Star Binaries}
\label{NumSim}
In this section, we discuss the numerical techniques and the set up of the BNS systems used in Bauswein \textit{et al.}~\cite{Bauswein:2020xlt} and K\"{o}lsch \textit{et al.}~\cite{Kolsch:2021lub} to study the impact of mass-ratio on the onset of a prompt-collapse to a BH.

The different groups use independent NR codes on similar binaries for their analysis.
The NR simulations of K\"{o}lsch \textit{et al.} use \texttt{BAM} code for dynamical evolutions which is based on the 3+1 decomposition of Einstein's Equation in the Z4c formalism \cite{Bernuzzi:2009ex,Hilditch:2012fp}.
The ($1+\log$) slicing and gamma-driver gauge conditions are used for lapse and shift, respectively~\cite{van_Meter_2006,Alcubierre:2002kk,doi:10.1142/2569}.
\texttt{BAM} uses nested Cartesian grids to solve both the general relativistic hydrodynamics (GRHD) equations and the Einstein's equations.
On the other hand, the smoothed particle hydrodynamics (\texttt{SPH}) code~\cite{Oechslin:2001km,Oechslin:2006uk,Bauswein:2010dn} used by Bauswein \textit{et al.} employs the conformally flat approximation of general relativity to solve Einstein's equations~\cite{CFC,Wilson:1996ty}.
The \texttt{SPH} code uses the Lagrangian particle method to solve the equations of GRHD~\cite{doi:10.1146/annurev.aa.30.090192.002551} and the multigrid solver to solve the field equations.
The particle mesh codes are then used to translate between the \texttt{SPH} particles and the gravity grid~\cite{CSP}.
\texttt{BAM} utilises AMR (adaptive mesh refinement) techniques~\cite{BERGER1984484} whereas the multilevel adaptive technique~\cite{Rhess} is used in \texttt{SPH} for solving nested grid setup.

For their analysis, K\"{o}lsch \textit{et al.} consider 290 NR simulations with varying total mass, and mass-ratios $q$, ranging from $0.57 \leqslant q \leqslant 1$.
They consider a restricted set of 3 EOSs, namely ALF2 \cite{Alford:2004pf}, SLy \cite{Douchin:2001sv}, and H4 \cite{Lackey:2005tk} such that their maximum mass is compatible with the observation of binary pulsar J0348+0432~\cite{Antoniadis:2013pzd}.
The maximum mass of non-rotating NSs corresponding to these EOSs are 1.99 $M_{\odot}$, 2.06 $M_{\odot}$ and 2.03 $M_{\odot}$, respectively.
However, as mentioned in K\"{o}lsch \textit{et al. } due to a small set of EOSs, the $M_{\rm thr}$ fit based only on their data does not reliably predict $M_{\rm thr}$ for other NR simulated EOSs given in Perego \textit{et al.}~\cite{Perego:2021mkd}.
  Therefore, as in K\"{o}lsch \textit{et al.} we would consider a fit derived from both the data \cite{Kolsch:2021lub,Kashyap:2021wzs,Perego:2021mkd} (Table X  of~\cite{Kolsch:2021lub}) for our analysis.
On the other hand, Bauswein \textit{et al.} consider a total of 40 EOS models, out of which 23 are hadronic EOS models {which they refer to as their ``base sample''}~\cite{Banik:2014qja,fortin2018,Marques:2017zju,Hempel:2009mc,Typel:2009sy,Typel:2005ba,Alvarez-Castillo:2016oln,Akmal:1998cf, Goriely:2010bm,Wiringa:1988tp,LATTIMER1991331,Shen:2011kr,Lalazissis:1996rd,Hempel:2011mk,Douchin:2001sv,Steiner:2012rk,Muther:1987xaa,Alford:2004pf,Engvik:1995gn,Schneider:2019vdm,Read:2008iy}, 8 belong to the ``excluded hadronic sample” of EOSs~\cite{LATTIMER1991331,Shen:2011kr,Lalazissis:1996rd,Hempel:2011mk, Lackey:2005tk,Glendenning:1984jr,Sugahara:1993wz,Toki:1995ya} and the remaining 9 are EOSs which include a phase transition to deconfined quark matter called `hybrid' EOS models~\cite{Bastian:2018wfl,Cierniak:2018aet,Fischer:2017lag,Bastian:2020unt,Bauswein:2018bma,Alvarez-Castillo:2016oln,Kaltenborn:2017hus}.
They consider three mass-ratios, namely 1.0, 0.85 and 0.7.
Based on the classification of the EOS models, Bauswein \textit{et al.} present the threshold mass fit formula for each EOS model as well as for the combinations of EOS models.
As the EOSs considered in K\"{o}lsch \textit{et al.} and Perego \textit{et al.} are included in the base and excluded EOS models of Bauswein \textit{et al.}, we only consider the threshold mass fit formula for the ``base+excluded'' model for our analysis on GW170817.

To verify the robustness of the two $q$-dependent threshold mass fits presented in Bauswein \textit{et al.} and K\"{o}lsch \textit{et al}, we use these fits on the 17 NR studies considered in Agathos \textit{et al.}~\cite{Agathos:2019sah}.
In that study, the authors used previously presented NR simulations data from Refs.~\cite{Radice:2018xqa,Dietrich:2018phi,Radice:2018pdn,Bernuzzi:2014owa,Radice:2017lry,Radice:2016rys,Perego:2019adq} along with the two NR simulations performed on the \texttt{WhiskyTHC} code~\cite{Radice:2012cu,Radice:2013xpa,Radice:2013hxh}.
The study considers mass-ratios ranging from $0.86 \leqslant q \leqslant 1.0$ with a total of 8 EOSs, out of which 5 are microphysical EOSs: BHB$\Lambda\phi$~\cite{Banik:2014qja}, DD2~\cite{Hempel:2009mc,Typel:2009sy}, LS220~\cite{LATTIMER1991331}, SFHo~\cite{Steiner:2012rk}, SLy-SOR~\cite{Schneider:2017tfi} and the remaining 3 are piecewise polytropic EOSs: ALF2, 2B and SLy~\cite{Read:2008iy}.


 

\section[Methods]{Methods}
\label{sec:methods}

In this Section we detail the methods used to i.) establish numerical fits relating the progenitor binary parameters to the nature of the merger remnant and ii.) perform Bayesian inference on the GW data from a BNS signal in order to recover an improved measurement on NS matter properties, with a focus on GW170817.

\subsection[NR fits]{Numerical fits for remnant classification}
\label{sec:nrfits}

The threshold mass ($M_{\rm thr}$) can be used to reliably classify the merger remnant in the NS binaries, that is, if the total mass $M$ of the NS binary is larger than the threshold mass, then the merger product is a prompt collapse to a BH.
It is convenient to express the threshold mass as
\begin{equation}
M_{\rm thr} = k_{\rm thr} M^{\rm TOV}_{\rm max}
\label{eq:mthr}~,
\end{equation}
where, $k_{\rm thr}$ is a coefficient greater than unity that depends on the EOS, the mass-ratio $q$ and the spins, and $M^{\rm TOV}_{\rm max}$ is the mass of the heaviest stable non-rotating NS supported by the EOS.
The validity of this expression was first established using large sets of NR hydrodynamical simulations~\cite{Hotokezaka:2011dh,Bauswein:2013jpa}.

For a sample of 12 hadronic EOSs and equal-mass binaries, Ref.~\cite{Bauswein:2013jpa} observed an approximate EOS-independent linear relation between $k_{\rm thr}$ and the maximum compactness of the non-rotating NS ($C_{\rm max}$).
For a fixed mass-ratio of $q=0.7$, the study of mass-ratio effects on the threshold mass indicated a monotonic relation between the two parameters~\cite{Bauswein:2020aag}.
A fitting formula for $M_{\rm thr}$ dependent on the EOS and the mass-ratio was first derived by Bauswein \textit{et al.}~\citep{Bauswein:2020xlt}, and was shortly followed by an independent study by K\"{o}lsch \textit{et al.}~\cite{Kolsch:2021lub}, who also pointed out that not all EOSs give a monotonically increasing threshold mass with increasing mass ratio.
In Ref.~\cite{Tootle:2021umi}, Tootle \textit{et al.} were the first to derive a fitting formula for the threshold mass that also includes dependence on the NS spins; we shall not consider the effect of spins in the current work.

For our analysis, we consider the $M_{\rm thr}\left(M_{\rm max}^{\rm TOV}, R_{\rm max}, q\right)$ fit from Bauswein \textit{et al.} (``\Bplus'' from here on), which we denote by $M^{B}_{\rm thr}$ and show explicit expression in Eq.~\eqref{eq:bausweinfit},
and the $M_{\rm thr}\left(M_{\rm max}^{\rm TOV}, R_{\rm 1.6 M_{\odot}}, q\right)$ fit from K\"{o}lsch \textit{et al.} (``\Kplus'' from here on), which we denote by $M^K_{\rm thr}$ and show explicit expression in Eq.~\eqref{eq:kolschfit}.
In the above equations, $M_{\rm max}^{\rm TOV}$, $R_{\rm max}$, $R_{1.6 M_{\odot}}$ are maximum mass, radius at the maximum mass and radius at $1.6~M_{\odot}$ of the non-rotating NS, respectively.
We use the \Bplus and \Kplus numerical fits of $M_{\rm thr}$, as they better capture their NR results, with the exception of a more fine-tuned fit by K\"{o}lsch \textit{et al.} which instead of some fiducial radius, depends on the estimated value of threshold mass at $q=1$ (i.e. $M_{\rm thr}^{q=1}$) based on the NR simulations.
However, as $M_{\rm thr}^{q=1}$ is a quantity that can only be estimated by NR simulations for a given EOS, we cannot use this fit for our GW170817 analysis, where the EOS is being sampled over; we can only use it in our NR injection studies of Agathos \textit{et al.}~\cite{Agathos:2019sah} (see Sec.~\ref{sec:injections}).
We denote this fit by $M^{K1}_{\rm thr}$ and show its explicit expression in Eq.~\eqref{eq:kolsch1fit}.
\begin{widetext}
\begin{eqnarray}
\label{eq:bausweinfit}
M^{B}_{\rm thr}(M_{\rm max}^{\rm TOV}, R_{\rm max},q)  & = & -0.07794 - 5.011 (1-q)^3 + \left[0.1863 -1.970 (1-q)^3\right]R_{\rm max} 
\nonumber \\ & {} & + \left[0.497 + 11.098 (1-q)^3\right]M_{\rm max}^{\rm TOV},  \\
\label{eq:kolschfit}
M^K_{\rm thr}(M_{\rm max}^{\rm TOV}, R_{\rm 1.6 M_{\odot}},q)  & =  & 0.246+\left[ 0.463+ 0.735 (1-q) +0.172 (1-q)^3\right]M_{\rm max}^{\rm TOV} \nonumber \\ & {} & + \left[0.141  - 0.116(1-q) - 0.214(1-q)^3 \right]R_{1.6 M_{\odot}},\\
\label{eq:kolsch1fit}
M^{K1}_{\rm thr}(M_{\rm thr}^{q=1}, M_{\rm max}^{\rm TOV}, q) & = & 0.281-\left[ 0.03667-1.11(1-q)-1.191(1-q)^3\right]M_{\rm max}^{\rm TOV} \nonumber \\ & {} & + \left[0.932 -0.753 (1-q)-1.625(1-q)^3\right]M_{\rm thr}^{q=1}~.
\end{eqnarray}
\end{widetext}
We will also compare the predictions of the $q$-dependent $M_{\rm thr}$ fits listed above with the $q$-independent $M_{\rm thr}$ fits used in Agathos \textit{et al.}. The latter were derived as a relation between $k_{\rm thr}$ and $C_{\rm max}$,
\begin{equation}
k_{\rm thr}(C_{\rm max}) = - (3.29 \pm 0.23)C_{\rm max} + (2.392 \pm 0.064)~,
\label{eq:Agathos}
\end{equation}
by combining the results of Refs.~\cite{Bauswein:2013jpa, Koppel:2019pys} with data by the \texttt{CoRe} collaboration \cite{Zappa:2017xba,Dietrich:2018phi}.


\subsection{Bayesian analysis of neutron star binaries}
\label{sec:bns-pe}
We perform Bayesian analysis on the GW detector data using a waveform model with matter effects that depend on the masses and the EOSs, to recover posterior samples of compact binary coalescence (CBC) parameters of the signal GW170817.
In addition to the standard CBC parameters for BBH---masses, spins, distance, coalescence time, phase, sky location and orientation angles---we also treat the EOS as an unknown function which we need to sample over.
We use the four-dimensional family of EOSs spanned by the spectral parametrization~\cite{Lindblom:2010bb,Carney:2018sdv,LIGOScientific:2018cki}, where the adiabatic index $\Gamma$ is approximated by a third-order polynomial $\Gamma(\rho) = \exp\left(\Sigma_{k=0}^{3}\gamma_k \log(p/p_0) \right)$.
We therefore sample over the additional parameters ($\gamma_0$, $\gamma_1$, $\gamma_2$, $\gamma_3$), to construct the EOS for the NS matter and derive all EoS-dependent parameters, such as tidal deformabilities, NS radii, etc. by solving the TOV equations for the sampled values of $m_1$ and $m_2$.

An obvious benefit of using this method is that for each point in the parameter space of ($\gamma_0$,$\gamma_1$, $\gamma_2$, $\gamma_3$), we can also calculate global properties of the EOS function like $M_{\rm max }$, $R_{\rm max}$, and $R_{\rm 1.6M_{\odot}}$.
We can then use these in our fits (Eqs.~\eqref{eq:bausweinfit},\eqref{eq:kolschfit}), to compute the threshold mass $M_{\rm thr}$ for each sample and thus translate the joint posterior of $(m_1, m_2, \gamma_0, \gamma_1, \gamma_2, \gamma_3)$ to the joint posterior probability of $(M_{\rm thr},M)$, where $M$ is the total mass of BNS system.
%

Finally, we classify each sample to either a formation of a prompt collapse to a BH or a NS remnant (non-prompt collapse), based on whether $M$ is greater or less than the threshold mass.
As the observation of EM counterparts after the observation of the GW170817 signal indicates that it was most likely a non-prompt collapse signal, we can reject a posteriori the part of the BNS parameter space for which we predict a prompt collapse.
We can then compare the resulting posteriors with the original results from the LIGO-Virgo analyses~\cite{LIGOScientific:2018hze,LIGOScientific:2018cki}, and assess the effect of our truncating procedure on measuring the properties of the BNS source.

We repeat the analysis described above by imposing a hard lower bound on the maximum nonrotating TOV mass of the NS EOS, to be consistent with the observation of the binary pulsar J0348+0432~\cite{Antoniadis:2013pzd}.
The measured mass of the pulsar is $M_{J0348+0432} = 2.01 \pm 0.04 \Ms$, so we impose a conservative constraint on the maximum TOV mass as $M_{\rm max}^{\rm TOV} \geq 1.97 M_{\odot}$.
This effectively removes a non-negligible part of the EOS parameter space, in particular the subset of models that are too soft to support NSs at masses larger than $1.97 \Ms$.
In Sec.~\ref{sec:results}, we will compare the improvement in parameter estimation with and without this constraint.

The {\PhenomPNRT} waveform model was used to measure the source parameters of GW170817 for both the parameter estimation studies of Ref.~\cite{LIGOScientific:2018hze} and the EOS studies of Ref.~\cite{LIGOScientific:2018cki}.
The BBH baseline in {\PhenomPNRT} model is based on the point-particle model presented in Ref.~\cite{Hannam:2013oca}, which incorporates effective-one-body (EOB) and NR-tuned tidal effects~\cite{Dietrich:2017aum,Dietrich:2018uni}, spin-induced quadrupole effects~\cite{PhysRevD.57.5287, Arun:2008kb,Mikoczi:2005dn,Bohe:2015ana,Mishra:2016whh} and precession effects.
Regarding the matter sector, Ref.~\cite{LIGOScientific:2018cki} makes minimal assumptions about the nature of the source and samples over the tidal deformabiltity parameters in contrast to Ref.~\cite{LIGOScientific:2018hze} where the condition that both NSs in the binary are described by the same EOS is imposed and the 4-dimensional spectral decomposition is used to sample the space of EOSs.
The two studies of Refs.~\cite{LIGOScientific:2018cki,LIGOScientific:2018hze} also compare the results between different waveform models, namely {\SEOBNRT}\cite{Bohe:2016gbl}, {\PhenomDNRT}~\cite{Husa:2015iqa} and {\TF} and found that the systematic uncertainties due to different waveform models are less than the statistical uncertainties in observing GW170817 signal.

\subsection{Validation of threshold mass fits}
\label{sec:bns-pe}
In addition to the GW170817 re-analysis, we implement the above Bayesian analysis on the injection study of the BNS systems given in Agathos \textit{et al.}~\cite{Agathos:2019sah}.
From the obtained joint posterior probability of $(M_{\rm thr}, M)$, we classify merger outcome into either prompt-collapse to BH or non-prompt collapse, and compute the posterior probability of prompt-collapse, given by
\begin{equation}
  P_{\rm PC} = P (M > M_{\rm thr}|d, \mathcal{H}_{\rm PC}) ~,
\end{equation}
for each BNS system of Agathos \textit{et al.}.
Here, by $d$ we denote the simulated data of the detector network (in this case from LIGO Livingston (L1), LIGO Hanford (H1) and Virgo (V1) at design sensitivity), and by $\mathcal{H}_{\rm PC}$ we denote the underlying model that we use for estimating the prompt collapse threshold for the total mass.
We then also run our above analyses with the additional maximum mass constraint from J0348+0432 and observe the effect this has on the estimate of $P_{\rm PC}$.

The waveform model used for the BNS injection studies of Agathos \textit{et al.} is \texttt{TEOBResumS}~\cite{Bernuzzi:2014owa,Nagar:2018zoe} with the non-spinning tidal model of Ref.~\cite{Akcay:2018yyh} and source parameters matching those of the NR simulations, as NR data do not span sufficiently many inspiral cycles to be used for injections.
The Bayesian data analysis to recover the source parameters is performed using {\TF} and {\PhenomPNRT}, restricting to the low-spin prior of~\cite{LIGOScientific:2018hze}.
However, as mentioned in Agathos \textit{et al.}, due to an error in the \texttt{NRTidal} version used, the {\PhenomPNRT} analysis led to biases in the tidal deformability inference; we therefore perform our analysis only on their {\TF} results.

\section{Results}
\label{sec:results}
In this Section we discuss our results corresponding to i.) the analysis for the GW170817 event and ii.) the analysis for injection studies of Agathos \textit{et al.}~\cite{Agathos:2019sah}.
\subsection{GW170817 Analysis}
\label{sec:results_gw170817}
We assume that the remnant of GW170817 did not promptly collapse to a black hole and apply our method of Sec.~\ref{sec:bns-pe} at the post-processing level, to the posteriors of the spectral EOS parameters for the two analyses of the GW170817 signal performed in Ref.~\cite{LIGOScientific:2018cki}:
\begin{itemize}
\item without maximum mass constraint on $M_{\rm max}^{\rm TOV}$
\item with the hard constraint on the maximum mass, $M_{\rm max}^{\rm TOV} > 1.97 M_{\odot}$, consistent with J0348+0432~\cite{Antoniadis:2013pzd}
\end{itemize}

We use the two sets of q-dependent $M_{\rm thr}$ fits of \Bplus and \Kplus\negthinspace.
For the \Kplus fits, however, we only present the analysis for the maximum mass constraint data sample, as the best $M_{\rm thr}$ fit given in Ref.~\cite{Kolsch:2021lub} is based on the NS radius at $1.6M_{\odot}$, i.e. $R_{1.6 M_{\odot}}$~\footnote{That fiducial NS mass is larger than the maximum mass supported by some samples in the unconstrained dataset.
Nevertheless, we confirm that after truncating these samples, the resulting posteriors agree well with those derived from the constrained dataset.
}.

\begin{figure*}
\begin{center}
\subfigure[]{\label{subfig:1a}\includegraphics[width=0.55\textwidth]{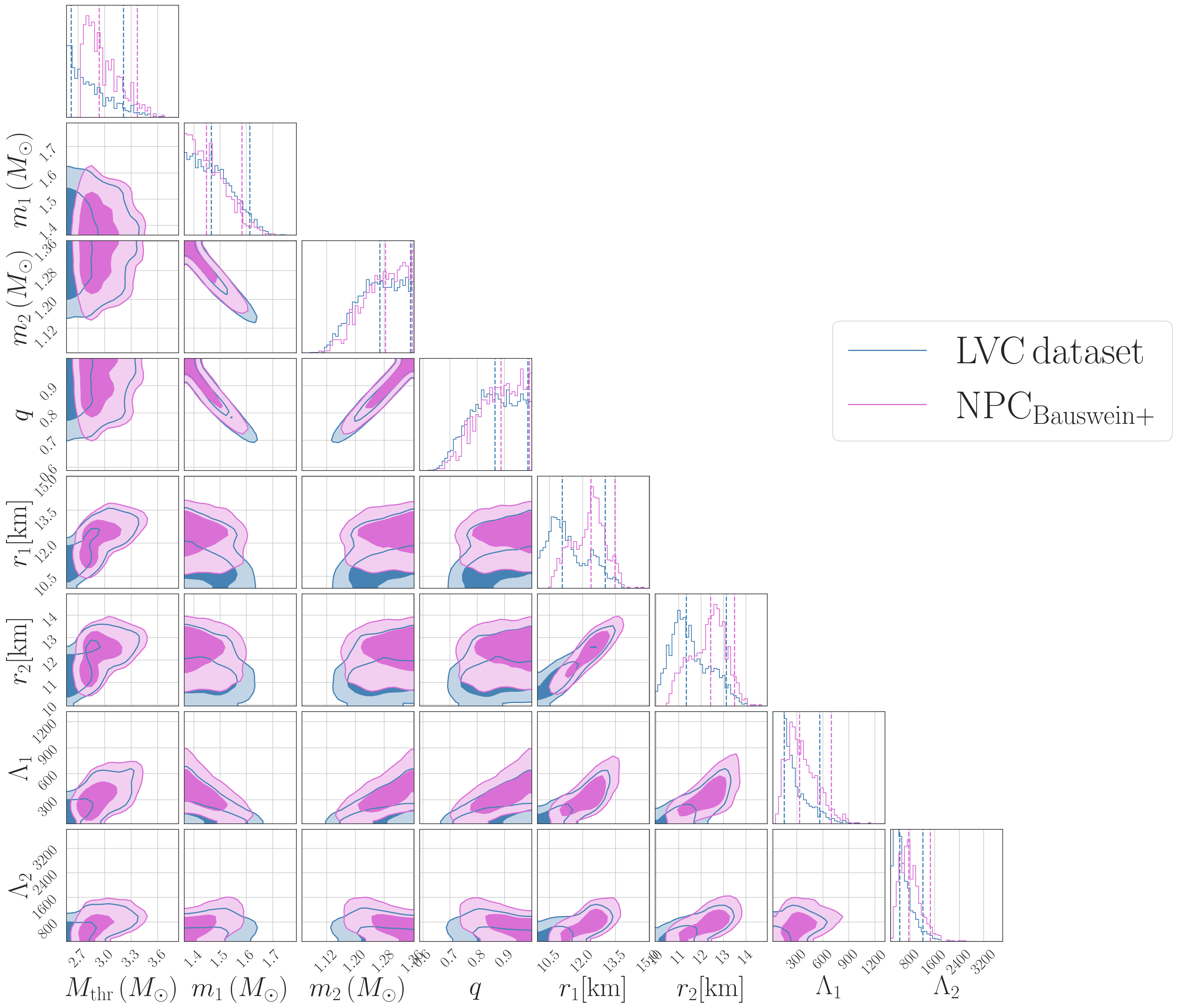}}
\hskip 1.cm \subfigure[]{\label{subfig:1b}\includegraphics[width=0.55\textwidth]{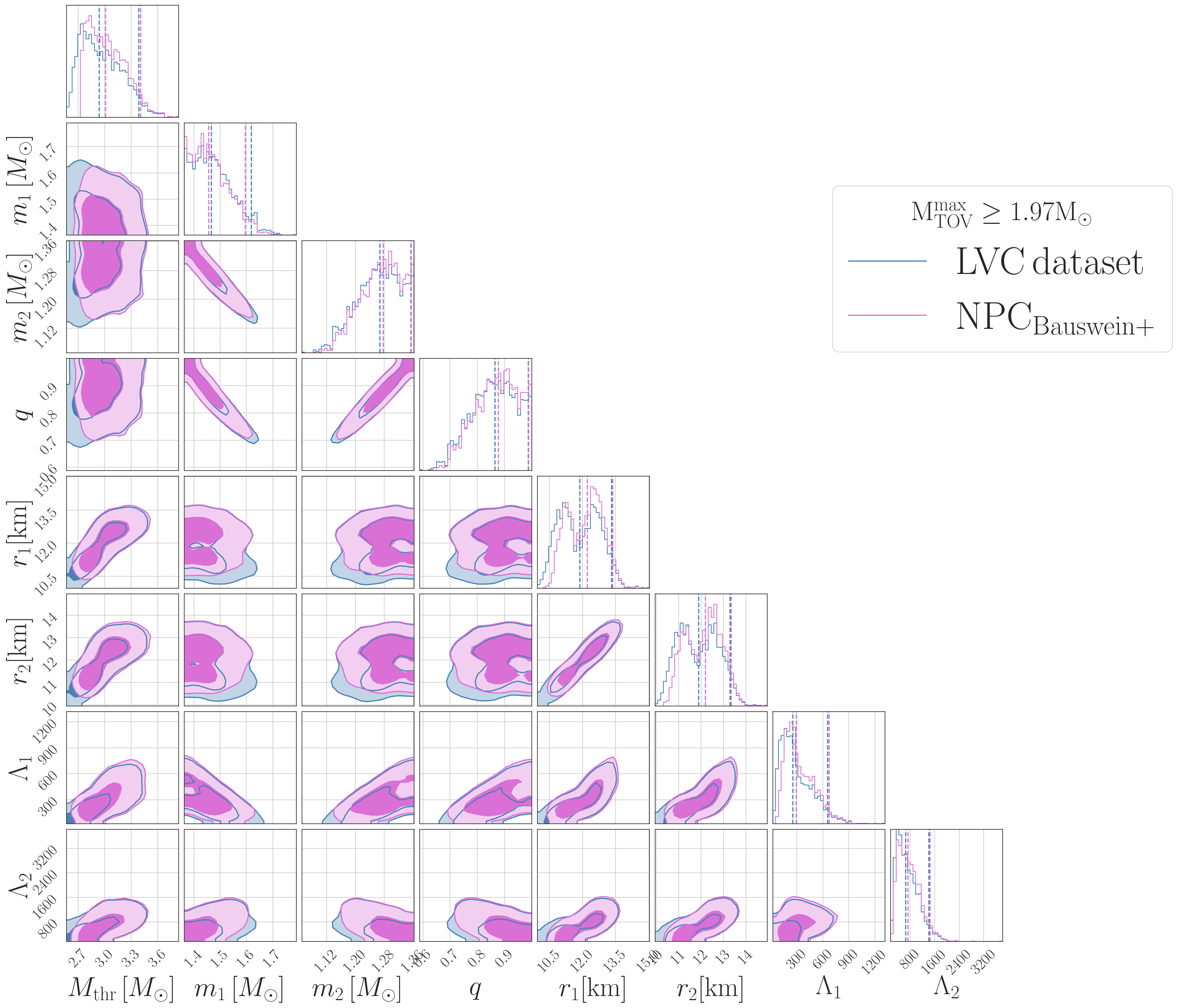}}
\end{center}
\caption {Corner plot for $(M_{\rm thr}, m_1, m_2, \Lambda_1, \Lambda_2, q, r_1, r_2)$ of the GW signal GW1701817 based on the \Bplus fit. Blue and purple colors indicate the LIGO-Virgo (LVC) and the corresponding non-prompt collapse (NPC) data samples, respectively. Contours correspond to the 50\% and 90\% confidence regions for the 2D joint posteriors, whereas in the 1D marginalized posteriors dotted lines show the 90\% credible interval for each parameter. The plots correspond to (a): the data sample without maximum mass constraint, and (b): the data sample with a hard lower bound on maximum mass at $1.97 M_\odot$.}
\label{fig:1}
\end{figure*}

\begin{figure*}
\begin{center}
\includegraphics[width=0.56\textwidth]{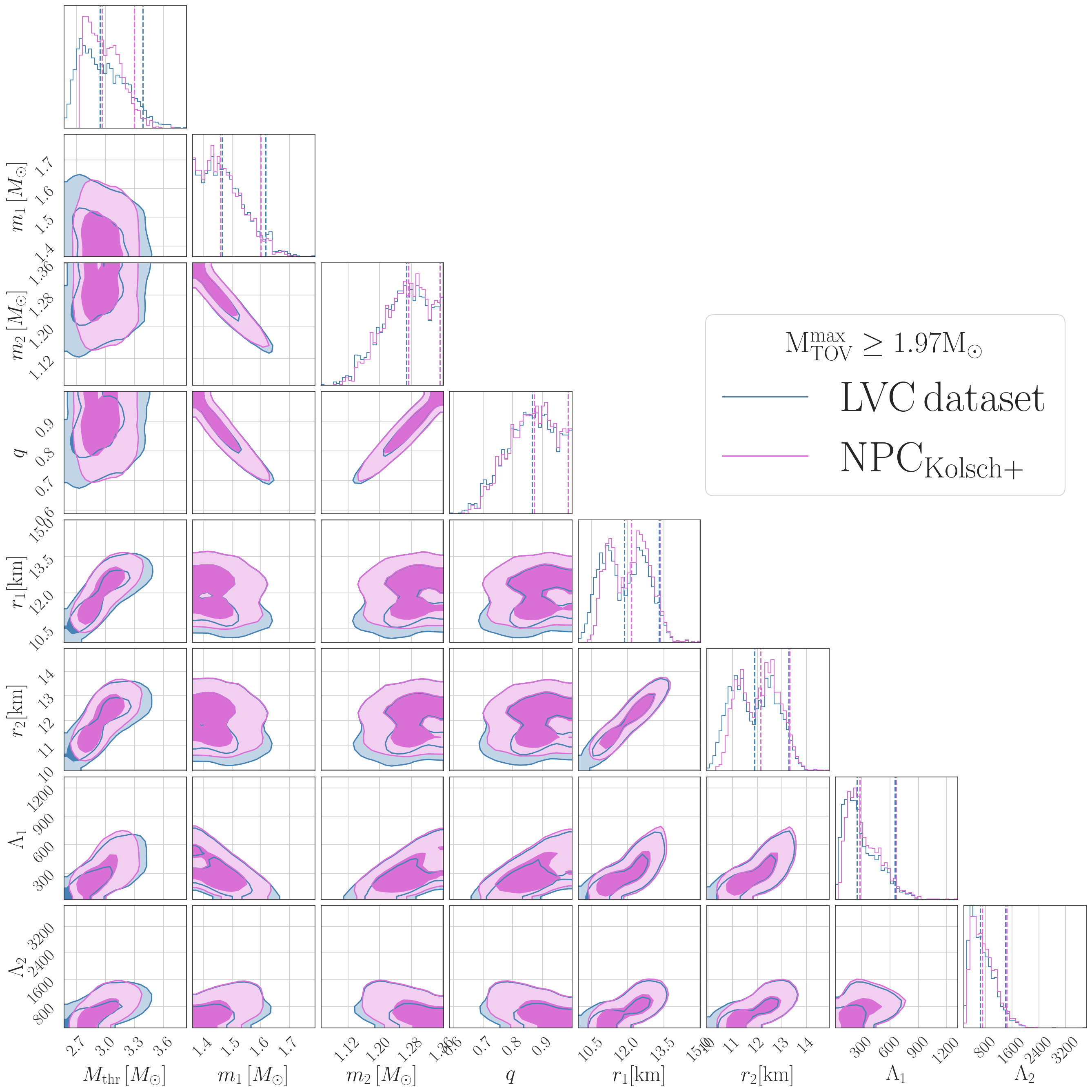}
\end{center}
\caption {Same as Fig.~\ref{fig:1}, but using the \Kplus fit; both datasets are generated using the hard lower bound on the maximum mass at $1.97 M_\odot$.}
\label{fig:2}
\end{figure*}

We compare the non-prompt collapse subset of the binary parameters like radii ($r_1$, $r_2$), (dimensionless) tidal deformabilities ($\Lambda_1$, $\Lambda_2$), mass ratio ($q$), and component masses ($m_1$,$m_2$) evaluated using the $M_{\rm thr}$ fits with the corresponding parameters of the complete LIGO-Virgo data sample.
The results of our analysis for GW170817 posteriors for both \Bplus and \Kplus fits are shown in Fig.~\ref{subfig:1a} and \ref{subfig:1b}, and Fig.~\ref{fig:2}, respectively.

Within the 90 \% credible limit, we obtain 
$m_1\in (1.37, 1.61) M_\odot$, $m_2\in(1.16,1.35)M_\odot$ for the unconstrained dataset, and  $m_1\in (1.37, 1.62) M_\odot$, $m_2\in(1.16,1.35)M_\odot$ for the constrained dataset, consistent with the results of Ref.~\cite{LIGOScientific:2018hze}.
After removing the prompt collapse cases using \Bplus fits {(64.80\% and 23.04\% of the data sample without the constraint and with the constraint, respectively)} from the data sample, we obtain the component masses for the remaining non-prompt collapse data sample as $m_1\in (1.37, 1.58) M_\odot$, $m_2\in(1.18,1.36)M_\odot$ for the unconstrained dataset, and $m_1\in (1.37, 1.60) M_\odot$, $m_2\in(1.17,1.35)M_\odot$ for the constrained dataset, which within the 90\% credible limit are also consistent with the results of Ref.~\cite{LIGOScientific:2018hze}.
Whereas, after removing the prompt collapse cases using the \Kplus fit (14.61\% of the data sample) from the constrained dataset, we obtain the component masses for the remaining non-prompt collapse data sample as $m_1\in (1.37, 1.59) M_\odot$ and $m_2\in(1.17,1.35) M_\odot$ are also consistent with the results of Ref.~\cite{LIGOScientific:2018hze} within the 90\% credible limit.

For the radii of the two NSs, within the 90 \% credible limit, we obtain $r_1 = 11.2^{+ 1.8}_{-1.5}$ km, $r_2 = 11.5^{+1.6}_{-1.3}$ km for the data sample without maximum mass constraint, and $r_1 = 11.9^{+1.4}_{-1.4}$ km, $r_2 = 11.9^{+1.4}_{-1.3}$ km for the data sample with maximum mass constrain \cite{Agathos:2019sah}.

Hence, we see that the maximum mass constraint rules out the possibility of radii $<10 {\rm km}$ compared to the data sample without maximum mass constraint.
This is due to the fact that the soft EOSs that predict low NS radii do not support massive NSs.
For the non-prompt collapse data sample based on the \Bplus fit, we obtain $r_1=12.3^{+1.2}_{-1.3}$ km, $r_2 = 12.3^{+1.2}_{-1.4}$ km for the unconstrained dataset and $r_1 = 12.1^{+1.3}_{-1.2}$ km, $r_2 = 12.1^{+1.2}_{-1.2}$ km for the constrained dataset.
This shows that under the non-prompt-collapse hypothesis, based on the $q$-dependent \Bplus fit, our GW analysis alone already rules out the possibility of NS radii  $< 10 {\rm km}$, even without using observations of heavy NSs.

Based on the \Kplus fit, the radii of the two NSs within 90\% credible limit for the non-prompt collapse dataset are now $r_1 = 12.1^{+1.2}_{-1.2}$~km and $r_2  = 12.1^{+1.2}_{-1.3}$~km.
The \Kplus fit also rules out radii $< 10 {\rm km}$ consistent with the findings of Refs.~\cite{Agathos:2019sah,LIGOScientific:2018hze} for the constrained dataset, as well as with the $R_{1.6}$ bounnd of Ref.~\cite{Bauswein:2017vtn}.
.
Additionally, we observe an improvement in the radii constraints as the uncertainty of NS radii decreases for the non-prompt collapse data samples.
The uncertainty in NS radii for the unconstrained LVC dataset decreases from 3.3 km to 2.6 km after removing the prompt collapse cases using the \Bplus fit. Similarly, the uncertainty in NS radii for the constrained LVC dataset decreases from 2.8 km to 2.5 km for \Bplus and \Kplus fits, based on non-prompt-collapse data samples.
This improvement in constraints on the radii implies improved constraints on the EOSs of the NS.

\begin{figure*}
\begin{center}
\subfigure[]{\label{subfig:3a}\includegraphics[width=0.75\textwidth]{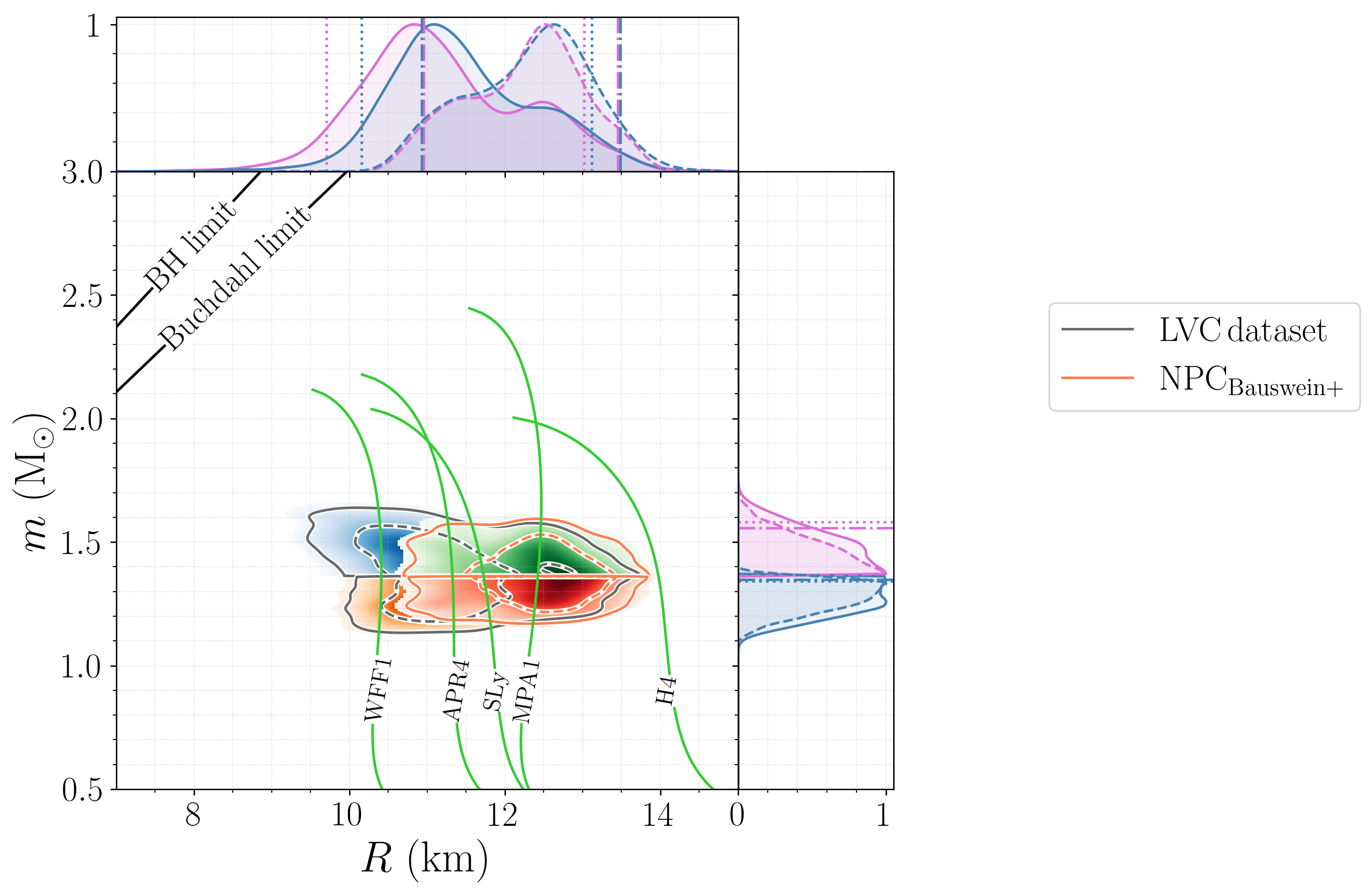}}
\hskip 1.cm \subfigure[]{\label{subfig:3b}\includegraphics[width=0.75\textwidth]{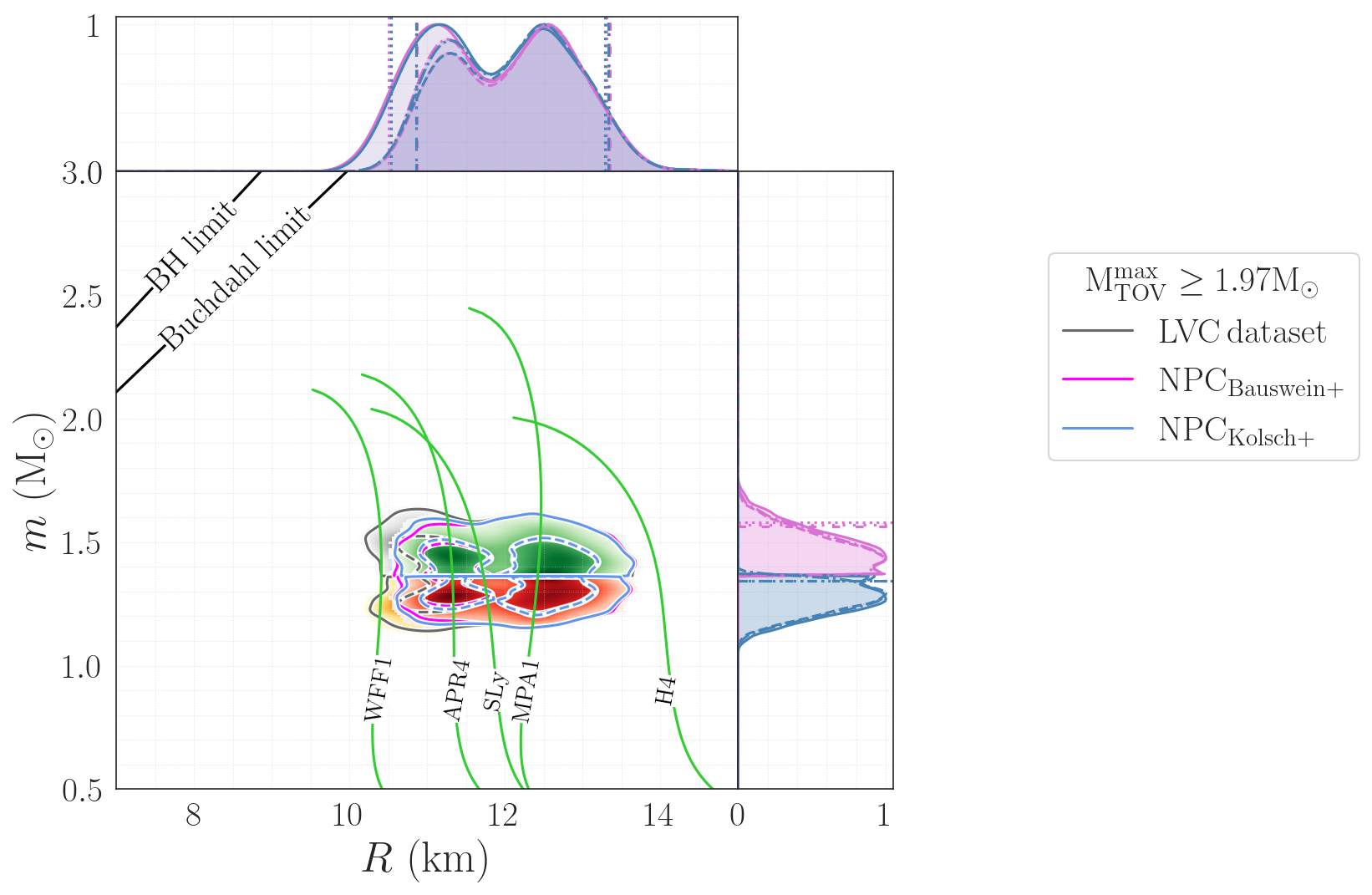}}
\end{center}
\caption {Mass-Radius posterior plots for the mass `$m$' and radius `$R$' of each binary component of GW signal GW170817 for both (a): the data sample without the maximum mass constraint, and (b): the data sample with the hard constraint on maximum mass, $M_{\rm max}^{\rm TOV} > 1.97 M_\odot$, corresponding to observation of binary pulsar J0348+0432. The black solid lines in top left indicates the Schwarzschild limit ($R=2m$), and the Buchdahl limit($R=\frac{9}{4}m$). Mass-Radius curves of selected EOSs are plotted in green color. In the one-dimensional plots, solid and dotted lines indicate the LIGO-Virgo (LVC) data sample and its 90 \% credible limit bounds, respectively, and dashed and dashdotted lines indicate the corresponding non-prompt collapse (NPC) data sample and its the 90 \% credible limit bounds,  respectively. }
\label{fig:mass-radius}
\end{figure*}

To illustrate the impact of our analysis, we plot the mass-radius curve for the posterior of the GW170817 signal along with the mass-radius curve for the selected EOSs (WFF1, APR4, SLy, MPA1, H4) in Figs.~\ref{subfig:3a} and \ref{subfig:3b}.
Within the 90\% credible limit, the EOSs consistent with the NPC dataset are only APR4, SLy, and MPA1, compared to the complete LVC dataset that was also supporting WFF1.

This follows from the improvement in the NS radii measurements, both in the case of the unconstrained dataset using the \Bplus fit for $M_{\rm thr}$, and in the cases of the constrained dataset using either the \Bplus fit or the \Kplus fit.

Overall, we observe that the effect of using a $q$-dependent classification of the remnant improves our contraints on the NS EOS and radii in much the same way as a maximum mass constraint does, i.e. by excluding part of the soft-EOS corner of the parameter space.
In this particular case, the effect of using the non-prompt-collapse hypothesis is stronger than the effect of imposing a maximum mass constraint at $1.97 \Ms$, as highlighted in Fig.~\ref{fig:mass-radius}.

\subsection{Injection studies}
\label{sec:injections}
To check the robustness of the \Bplus and \Kplus fits, we use progenitor and remnant information from the existing NR simulations of BNS mergers given in Agathos \textit{et al.}~\cite{Agathos:2019sah}.
We generate posteriors of CBC parameters using the {\TF} waveform model for each of these BNS systems.
Similar to the GW1701817 analysis of Sec.~\ref{sec:results_gw170817}, we consider the two cases of the BNSs, i.e. constrained and unconstrained datasets with respect to the maximum NS mass.
After classifying the merger remnant of each dataset using the above fits for the threshold mass, we calculate the cumulative probability distribution of prompt collapse, $P(M > M_{\rm thr})$.

\begin{figure*}
\begin{center}
\subfigure[]{\label{subfig:4a}\includegraphics[width=0.48\textwidth]{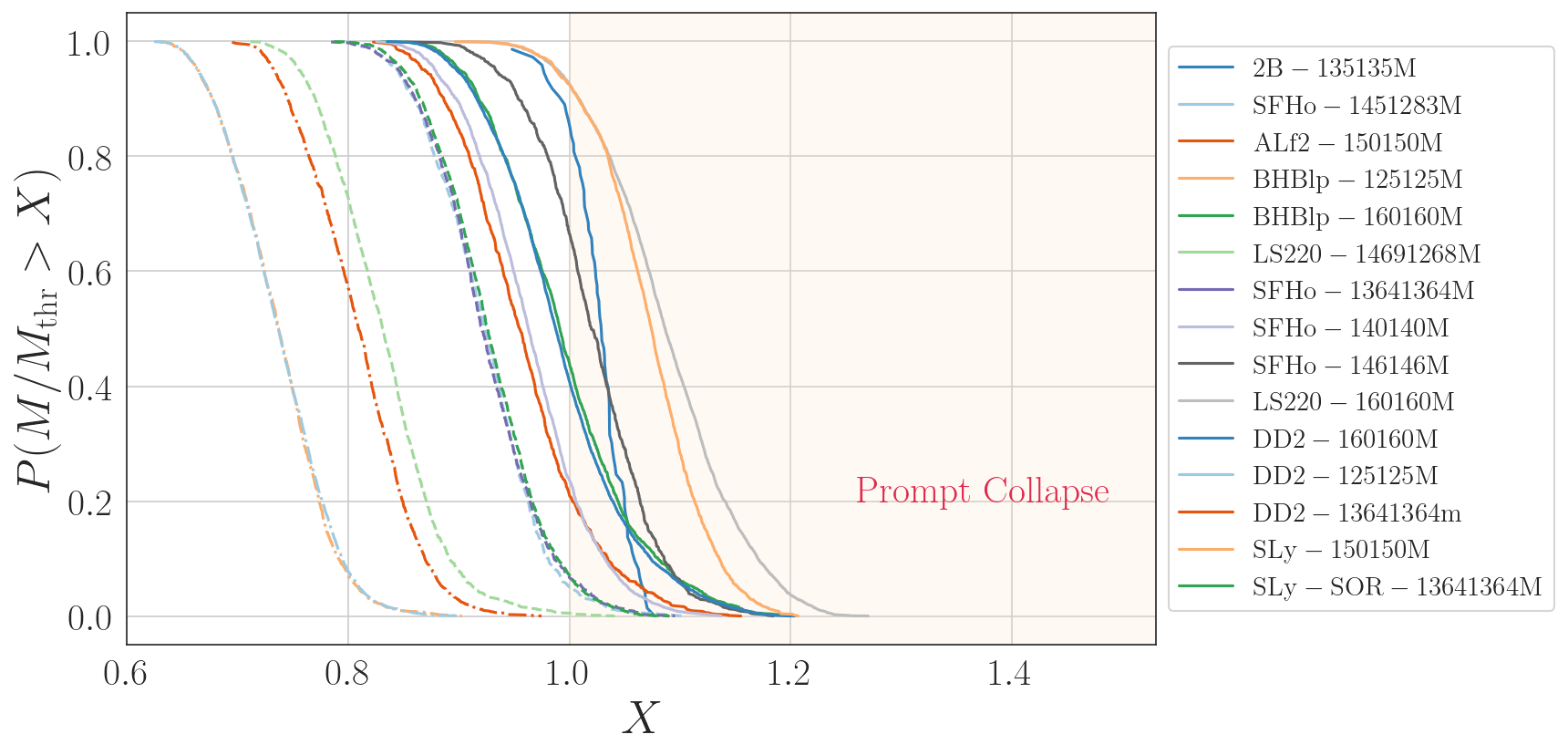}}
\subfigure[]{\label{subfig:4b}\includegraphics[width=0.48\textwidth]{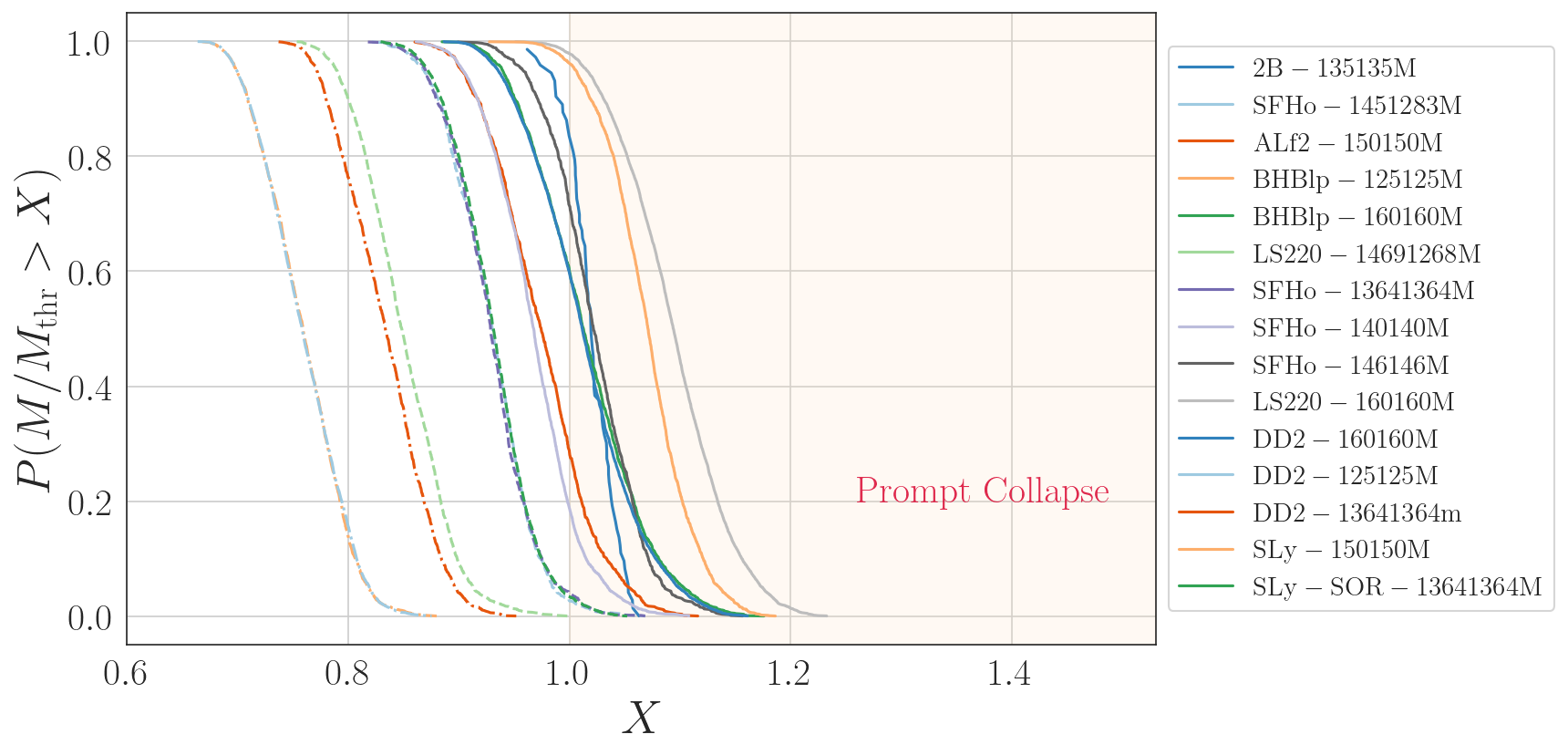}}
\end{center}
\caption {Cumulative probability of $M/M_{\rm thr}$, the ratio between total mass ($M$) and the threshold mass ($M_{\rm thr}$), for the injection study using the simulated BNS systems of Table~\ref{tab:tab1} with a hard lower bound on maximum mass at $1.97\Ms$. The probability of prompt collapse for each binary neutron star system is at $X=1$. The solid, dashed and dashed-dotted lines correspond to formation of a BH, HMNS, and MNS, respectively. The plots correspond to (a): \Bplus fit of Eq. (\ref{eq:bausweinfit}), and (b): \Kplus  fit of Eq. (\ref{eq:kolschfit}), respectively. }
\label{fig:4eos_test}
\end{figure*}

Here too, for the \Kplus fit we only calculate the cumulative probability of prompt collapse for the constrained dataset.
The cumulative probability distribution (CDF) plots for the constrained dataset using the \Bplus and \Kplus fits are shown in Fig.~\ref{subfig:4a} and Fig.~\ref{subfig:4b}, respectively~\footnote{For the unconstrained dataset CDFs with the \Bplus fit, see Fig.~\ref{fig:5} in Appendix~\ref{app:A}.}.
The vertical line in the plot corresponds to the threshold of prompt collapse (beyond which the total mass exceeds the threshold mass).
We observe that the $q$-dependent \Bplus and \Kplus fits correctly infer the merger outcome for most of the BNS models except for a few cases, all of which were also misclassified in Agathos \textit{et al.}.

Then we compare the probability of prompt collapse predicted by the $q$-dependent \Bplus and \Kplus fits with the one by the $q$-independent $M_{\rm thr}~$ fit of Agathos \textit{et al.} (see, Eq.~\eqref{eq:Agathos} for the fit formula).
The summary of the results based on these fit formulae along with the binary parameters and the outcome of the merger for each BNS is shown in Table \ref{tab:tab1}.
We observe that both \Bplus and \Kplus fits better capture the formation of prompt-collapse to a BH as the probability for prompt-collapse 
increases for both the cases of data samples compared to $q$-independent $M_{\rm thr}$ fit of Agathos \textit{et al.}.
Furthermore, we observe that compared to \Bplus, the \Kplus fit is better at predicting the type of merger remnant, as it typically gives higher probabilities when a prompt collapse does occur and lower probabilities in the cases resulting to a (meta-)stable remnant.
We see that even for the few cases where the merger remnant is misclassified (SFHo $1.4+1.4$, ALF2 $1.5+1.5$), with respect to the results of Agathos \textit{et al.}, the probability of prompt collapse has significantly moved towards the correct direction.

Finally, as mentioned in Sec.~\ref{sec:nrfits}, we also use the $M_{\rm thr}^{K1}$ fit~\cite{Kolsch:2021lub} dependent on the estimated value of the threshold mass at $(q=1)$ from NR simulation data (i.e., $M_{\rm thr}^{q=1}$), together with $M_{\rm max}^{\rm TOV}$ and $q$ for our analysis (see Eq.~\eqref{eq:kolsch1fit} for the fitting formula).
The values of the threshold mass at ($q=1$) for the EOSs of interest are obtained from the NR data of Refs.~\cite{Agathos:2019sah,Perego:2021mkd,Kolsch:2021lub,Kashyap:2021wzs} (see, Table~\ref{tab:tab2}).
The probability of prompt collapse computed corresponding to this fit is given in Table~\ref{tab:tab1}.
From Table~\ref{tab:tab1}, we see that this fit better captures both the formation of a prompt collapse to a BH and a non-prompt collapse compared to all other $M_{\rm thr}$ fits considered in our analysis for most of the BNS systems; however, it is only available for a small discrete set of EOSs.

\begin{table*}
\caption{\label{tab:tab1} Summary of probability of prompt collapse (\%) for the known inspiral-merger waveforms of Agathos\textit{et al.}~\cite{Agathos:2019sah}. The collapse time $t_{\rm BH}$ is time measured from the peak of GW amplitude, called the merger time. The threshold mass fit formulas $M_{\rm thr}(M_{\rm max}^{\rm TOV},C_{\rm max})$, $M_{\rm thr}(M_{\rm max}^{\rm TOV}, R_{\rm max}, q)$, $M_{\rm thr}(M_{\rm max}, R_{\rm 1.6 M_{\odot}}, q)$, and $M_{\rm thr}(M_{\rm max}^{\rm TOV}, M_{\rm thr, (q=1)},q)$ are given in Eqs.~\eqref{eq:bausweinfit}-\eqref{eq:Agathos}, respectively.}
\begin{ruledtabular}
\begin{tabular}{  c  c  c  c p{1.5cm} c c c c c c}
\textrm{EOS}&
\textrm{$m_1 (\rm M_{\odot})$}&
\textrm{$m_2 (\rm M_{\odot})$}&
\textrm{$t_{\rm BH}[ms]$}&
\textrm{Remnant at $t\sim 3ms$}&
\multicolumn{2}{r}{\textrm{\scalebox{.9}{$\begin{gathered}M_{\rm thr}(M_{\rm max}^{\rm TOV},C_{\rm max})\\ \textrm{Agathos~\textit{et al.}~\textrm{\cite{Agathos:2019sah}}}\end{gathered}$}}}&
\multicolumn{2}{r}{\textrm{\scalebox{.9}{$\begin{gathered}{M_{\rm thr}(M_{\rm max}^{\rm TOV}, R_{\rm max}, q)}\\ {\textrm{Bauswein}~\textit{et al.}~\textrm{\cite{Bauswein:2020xlt}}}\end{gathered} $}}}&
\multicolumn{1}{r}{\textrm{\scalebox{.9}{$\begin{gathered}\tiny{M_{\rm thr}(M_{\rm max}^{\rm TOV}, R_{\rm 1.6 M_{\odot}}, q)} \\ \textrm{K\"{o}lsch}~\textit{et al.}~\textrm{\cite{Kolsch:2021lub}}\end{gathered}$}}}&
\multicolumn{1}{c}{\textrm{\scalebox{.9}{$\begin{gathered}M_{\rm thr}(M_{\rm max}^{\rm TOV}, M_{\rm thr}^{q=1},q)\\ \textrm{K\"{o}lsch}~\textit{et al.}~\textrm{\cite{Kolsch:2021lub}}\end{gathered}$}}}\\
& & & & &\scalebox{.9}{$P_{\rm PC}^{\rm M_{\rm thr}}$}&\scalebox{.9}{$P_{\rm PC}^{\rm M_{\rm thr}, M_{\rm max}^{\rm TOV}}$}\hspace{0.5cm}&$P_{\rm PC}^{\rm M_{\rm thr}}$&$P_{\rm PC}^{\rm M_{\rm thr}, M_{\rm max}^{\rm TOV}}$&$P_{\rm PC}^{\rm M_{\rm thr}, M_{\rm max}^{\rm TOV}}$&$P_{\rm PC}^{\rm M_{\rm thr}, M_{\rm max}^{\rm TOV}}$\\
\colrule
\hline
2B & 1.35 & 1.35 & 0.49 &BH & 99.5 & 67.1&99.9 &86.3&83.6 & 100.0 \\ 

SFHo & 1.40 & 1.40 & 1.07 & BH & 46.7 & 8.0& 58.8 & 23.6& 18.8& 17.6 \\

SFHo & 1.46 & 1.46 & 0.70 &BH & 77.2& 48.1& 85.6 & 66.5&71.5&100.0 \\

ALF2&1.50&1.50&0.64&BH&25.7&5.7&36.6&21.1&28.6&84.0 \\

BHB$\Lambda\phi$ &1.60&1.60&0.99&BH&40.7&27.2&53.4&44.0&60.1&100.0 \\

LS220 & 1.60&1.60&0.63 & BH&92.4&88.1&95.3&92.6&97.9&100.0 \\

DD2&1.60&1.60&$\sim$3&BH&37.7&24.2&50.2&40.9&59.7&32.0 \\

SLy&1.50&1.50&0.99&BH&96.1&87.9&97.6&92.5&96.2&100.0 \\

SFHo & 1.364&1.364&$\sim$4&HMNS&25.7&0.5&34.1&6.7&4.0&3.4\\

SLy-SOR&1.364&1.364&$\sim$14&HMNS&29.8&0.6&39.1&7.2&3.4&4.8\\

SFHo & 1.45&1.283&$\sim$12&HMNS&25.9&0.4&34.7&5.0&2.7&2.9 \\

LS220 & 1.469&1.268&$\sim$33&HMNS&0.8&0.0&1.9&0.4&0.0&0.1\\

BHB$\Lambda\phi$ &1.25&1.25&$>$20&MNS&0.0&0.0&0.0&0.0&0.0&0.0 \\

DD2&1.25&1.25&$>$20&MNS&0.0&0.0&0.0&0.0&0.0&0.0 \\

DD2&1.364&1.364&$\sim$21&MNS&0.1&0.0&0.9&0.0&0.0&0.0 \\ \\ 
\end{tabular}
\end{ruledtabular}
\end{table*}

\begin{table}
\caption{\label{tab:tab2} NR simulation based values of $M_{\rm thr}^{q=1}$ for EOSs considered in Agathos \textit{et al.}~\cite{Agathos:2019sah}.
}
\begin{ruledtabular}
\begin{tabular}{  c  c  c }
\textrm{EOS}&
\textrm{$M_{\rm thr}^{q=1}$}&
\textrm{Reference}\\
\colrule
\hline
2B&2.43&\citep{Agathos:2019sah}\\
SFHo&2.824&\cite{Perego:2021mkd,Kashyap:2021wzs}\\
ALF2&2.963&\cite{Kolsch:2021lub}\\
BHB$\Lambda\Phi$&3.024&\cite{Perego:2021mkd,Kashyap:2021wzs}\\
LS220&2.956&\cite{Perego:2021mkd,Kashyap:2021wzs}\\
DD2&3.274&\cite{Perego:2021mkd,Kashyap:2021wzs}\\
SLy&2.756&\cite{Kolsch:2021lub}\\
\end{tabular}
\end{ruledtabular}
\end{table}
\section{Conclusions}
\label{sec:conclusions}
We analyse the gravitational wave signal GW170817 using the $q$-dependent threshold mass fits given in Bauswein \textit{et al.}~\cite{Bauswein:2020xlt} and K\"{o}lsch \textit{et al.}~\cite{Kolsch:2021lub}.
For our analysis, we consider the two cases of GW170817 signal - without the maximum mass constraint data sample and with the hard constraint on the maximum mass data sample corresponding to the observation of binary pulsar J0348+0432.
Using the threshold mass fits we identify the prompt collapse cases which we then remove from the two data samples for our analysis.
Then we compare the non-prompt collapse data sample of each case with its corresponding LVC data sample.
We observe that the non-prompt collapse data sample removes the radii $<$ 10 km even for without the maximum mass constraint data sample in comparison to results observed in Refs.~\cite{LIGOScientific:2018hze,Agathos:2019sah} where radii $<$ 10 km is not supported only by the data sample with the maximum mass constraint.
We also see that for the non-prompt collapse data sample the uncertainty of NS radii decreases implying improved constraints on the NS radii and hence the EOSs.
As an example, we observe that the EOS WFF1 is not supported by the non-prompt collapse data sample for both the cases of GW170817 signal compared to the results of Refs.~\cite{LIGOScientific:2018hze,Agathos:2019sah}.

The threshold mass fits are then validated with a set
of 17 injections given in Agathos \textit{et al.}~\cite{Agathos:2019sah}.
The merger remnant for all the signals are not just correctly inferred from both fits 
but 
the probability of the formation of a prompt collapse is also improved compared to the $q$-independent fit of Agathos \textit{et al.} with the exception of a few signals.
Additionally, we observe that the probability of formation of both a prompt collapse and a non-prompt collapse remnant has improved for $q$-dependent threshold mass fits of K\"{o}lsch \textit{et al.} compared to both the fits of Bauswein \textit{et al.} and Agathos \textit{et al.}.

Application of either of the fits, \Bplus{} or \Kplus{} to the GW170817 analysis gives similar improvements for the posterior PDFs of NS masses and radii.
This suggests that any differences betweent these two fits have relatively small impact for GW170817-like signals, that should also allow us to improve the constraints on the NS properties for similar detections in the future.
Nevertheless, the BNS parameter space, especially in the regions close to the prompt-collapse threshold, need to be explored more densely with future numerical relativity simulations, for a variety of EOSs.

\begin{acknowledgments}
The authors are grateful to Ulrich Sperhake, Sebastiano Bernuzzi and Alessandro Nagar for useful discussions during the preparation of this work and to Nikolaos Stergioulas for reviewing the manuscript and providing useful comments.
  T.J. is jointly funded by the Cambridge Trust, Department of Applied Mathematics and Theoretical Physics (DAMTP), and Centre for Doctoral Training, University of Cambridge. M.A. is supported by the Kavli Foundation.
  This material is based upon work supported by NSF's LIGO Laboratory which is a major facility fully funded by the National Science Foundation.
This work makes use of \texttt{GNU Scientific Library}, \texttt{NumPy}~\cite{Numpy}, \texttt{SciPy}~\cite{Scipy}, \texttt{Matplotlib}~\cite{Matplotlib}, \texttt{seaborn}~\cite{seaborn} and \texttt{LALsuite}~\cite{lal,lalsuite1} software packages.
\end{acknowledgments}

\appendix

\section{Effect of $M_{\rm max}^{\rm TOV}$ constraint}
\label{app:A}
\begin{figure}[H]
\begin{center}
\includegraphics[width=0.56\textwidth]{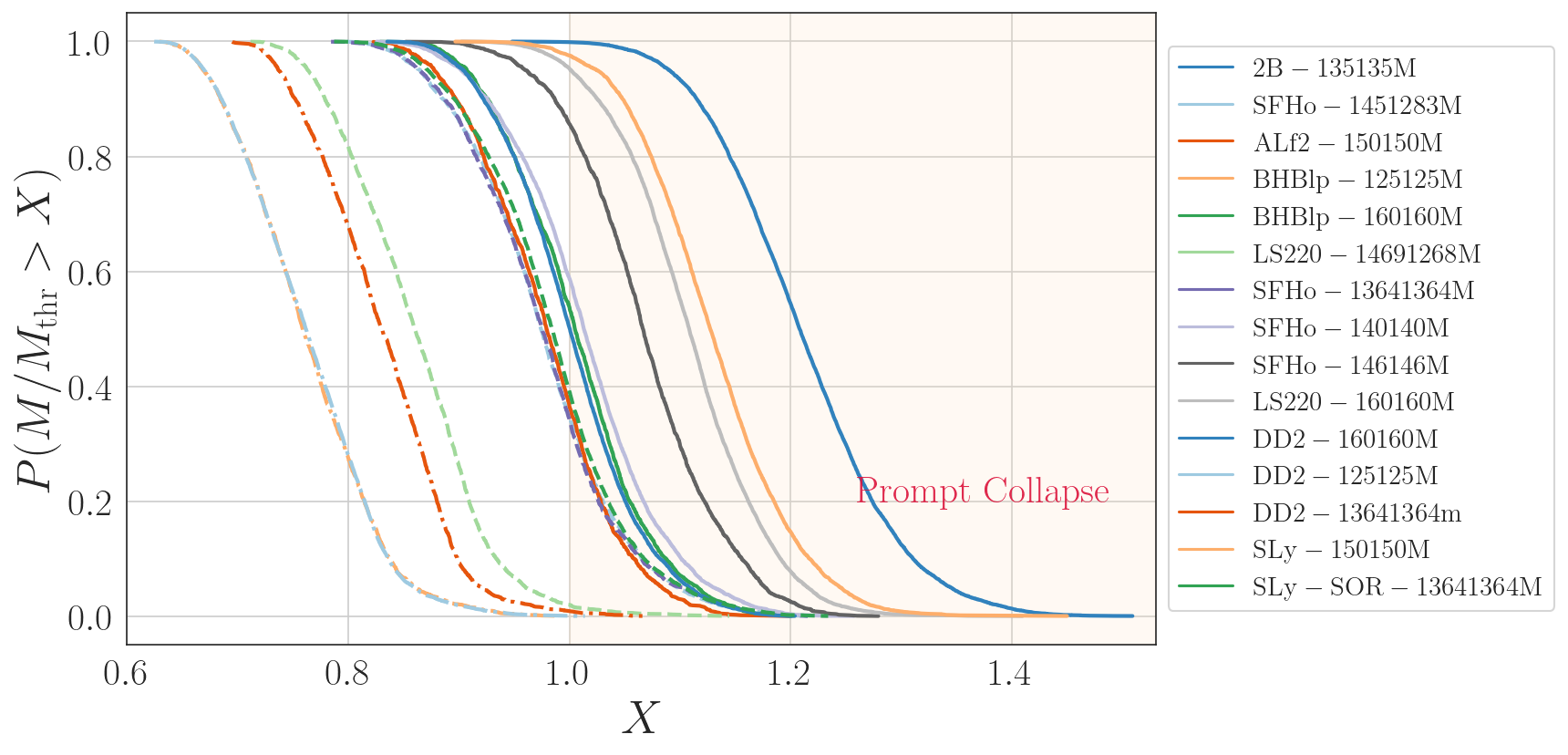}
\end{center}
\caption {Cumulative probability of $M/M_{\rm thr}$, the ratio between total mass ($M$) and the threshold mass ($M_{\rm thr}$), for the injection study using simulated BNS systems of Table~\ref{tab:tab1} for `\Bplus' fit of Eq.~\eqref{eq:bausweinfit}. The probability of prompt collapse for each binary neutron star system is at $X=1$. The solid, dashed and dashed-dotted lines correspond to formation of a BH, HMNS, and MNS, respectively.}
\label{fig:5}
\end{figure}

In this Appendix, we analyse the impact of the maximum mass constraint corresponding to the observation of binary pulsar J0348+0432 on the known BNS inspiral-merger waveforms of Agathos \textit{et al.}~\cite{Agathos:2019sah}. For this, we first calculate the cumulative probability of prompt collapse [$P(M>M_{\rm thr})$] for the unconstrained dataset and then compare it with the results presented in Fig.~\ref{subfig:4a}. The cumulative probability plot for the data sample without maximum mass constraint is shown in Fig.~\ref{fig:5}. 

On comparing Fig.~\ref{fig:5} with Fig.~\ref{subfig:4a}, we observe that ratio $M/M_{\rm thr}$ decreases for the constrained dataset compared to the unconstrained dataset. This can be due to increase in the $M_{\rm thr}$ values, which is expected, as the maximum mass constraint removes the soft part of the EOSs. 
This is also reflected from the results in Table~\ref{tab:tab1}, where we see a decreasing trend in the probability of prompt collapse for the constrained dataset compared to the unconstrained dataset.

\bibliography{references}

\end{document}